\documentclass[%
aps,
jmp,%
amsmath,amssymb,
reprint,%
]{revtex4-1}
\usepackage{graphicx}
\usepackage{multirow}
\usepackage{diagbox}
\usepackage{subfigure}
\usepackage{dcolumn}
\usepackage{bm}
\usepackage[mathlines]{lineno}
\usepackage{epstopdf}
\usepackage{amsmath}
\usepackage[colorlinks]{hyperref}
\hypersetup{
	colorlinks = true,
	citecolor = {blue},
	linkcolor = {red}
}

\newcommand{\blk}{\color{black}}

\begin{document}
	\title{High-Rate Free-Running Reference-Frame-Independent Measurement-Device-Independent Quantum Key Distribution with Classified Distillation}
	\author{Xin Liu$^{1}$, Zhicheng Luo$^{1}$, Kaibiao Qin$^{1}$, Jiawang Liu$^{1}$, Zhenrong Zhang$^{2}$, and Kejin Wei$^{1,*}$ }
	
	\address{
		$^1$Guangxi Key Laboratory for Relativistic Astrophysics, School of Physical Science and Technology,
		Guangxi University, Nanning 530004, China\\		$^2$Guangxi Key Laboratory of Multimedia Communications and Network Technology, School of Computer Electronics and Information, Guangxi University, Nanning 530004, China\\
		$^*$Corresponding author: kjwei@gxu.edu.cn
	}
	\date{\today}
	
	\begin{abstract}
    Reference-frame-independent measurement-device-independent quantum key distribution (RFI-MDI-QKD) eliminates detector side-channel attacks and avoids reference-frame calibration. While its feasibility has been widely demonstrated, existing implementations typically assume fixed or slowly drifting reference-frame misalignment, conditions rarely satisfied outside the laboratory. In realistic environments, rapid and free-running reference-frame variations can severely degrade both the key rate and transmission distance of conventional RFI-MDI-QKD.
    Here we propose a free-running RFI-MDI-QKD protocol that maintains high-rate key generation under rapid reference-frame variations. By introducing a classification-distillation method that reclassifies total detection events, secure keys can be extracted without modifying the experimental setup. Our protocol achieves a key rate more than nine times higher than the best previous RFI-MDI-QKD scheme and tolerates channel losses exceeding 24~dB, where earlier approaches fail. These results enable practical quantum key distribution on mobile platforms, including satellite-to-ground links and airborne nodes.

	\end{abstract}
	
	\maketitle
	
	\itshape Introduction.\upshape—Quantum key distribution (QKD) enables unconditionally secure communication grounded in information theory, with security guaranteed by the fundamental principles of quantum mechanics. Since the proposal of the BB84 protocol~\cite{BB84}, QKD has achieved remarkable advances in long-distance transmission~\cite{Lucamarini2018,Boaron2018,Wang2022,Zhou2023,Liu2023,Li2025microsatellite}, system miniaturization~\cite{Zhang2019integrated,Wei2023,Li2023,Du2024,Hajomer2024}, and real-world demonstrations~\cite{Avesani2022,Zhou2024,Zhu2024}. Meanwhile, commercial QKD systems are now available off the shelf and have been widely deployed~\cite{Chen2021integrated,Vest2022,Pittaluga2025}.

	Quantum key distribution (QKD) is theoretically secure; however, practical implementations are subject to security vulnerabilities arising from discrepancies between realistic devices and their idealized models~\cite{Gottesman2004,Gisin2006,Sun2015,Zhang2021,Chen2022,Huang2023,Ye2023,Trefilov2025,Gnanapandithan2025}. In particular, detectors in QKD systems are highly vulnerable to attacks, representing a critical security bottleneck~\cite{Qi2006,Lydersen2010}. 
	
	To address this limitation, the measurement-device-independent QKD (MDI-QKD) protocol was proposed, eliminating all detection-side channels~\cite{Lo2012,2012Braunstein}. Since its introduction, MDI-QKD has attracted extensive theoretical and experimental investigations~\cite{Wang2019Asymmetric,Wei2020,Woodward2021,Jiang2022,Li2023_Free-Space,Shao2025}. Numerous MDI-QKD variants have been developed to overcome technical challenges~\cite{2012Tmamki,2012Ma-time-bin-MDI,Yin2014}, increase transmission distances~\cite{Zeng2022,Xie2022}, or enhance security~\cite{Wang2022_side,Zhang2022_side}. 
	
	Building on MDI-QKD, the reference-frame-independent MDI-QKD (RFI-MDI-QKD) protocol, first introduced by Yin \textit{et al.}~\cite{Yin2014}, eliminates the need for reference-frame calibration during operation, substantially reducing system complexity. To date, the feasibility of RFI-MDI-QKD protocols has been thoroughly validated in both laboratory and field experiments~\cite{Wang2015,Liu2018,Liu2021RFIMDI,Zhu2023AD,Zhou2025}. Moreover, the feasibility of establishing MDI-QKD networks using RFI-MDI-QKD has also been demonstrated~\cite{Fan-Yuan2022}.

	A major limitation of previous RFI-MDI-QKD experiments is that they either assume a fixed reference-frame misalignment angle~\cite{Liu2018,Liu2021RFIMDI} or are tested only under slow in-laboratory reference-frame drift~\cite{Fan-Yuan2022}. In realistic setups, the reference frame is typically free-running and can vary rapidly due to channel disturbances caused by weather, temperature, and other environmental factors at the node locations. Such rapid variations can sharply reduce the key rate and transmission distance, or even cause the protocol to fail, resulting in no secure key bits being extracted. Consequently, the requirement for slow reference-frame drift severely limits the secret key rate and transmission distance of traditional RFI-MDI-QKD protocols, hindering their widespread deployment in quantum networks.
	
	\begin{figure}[!t]
		\centering
		\includegraphics[width=1\linewidth]{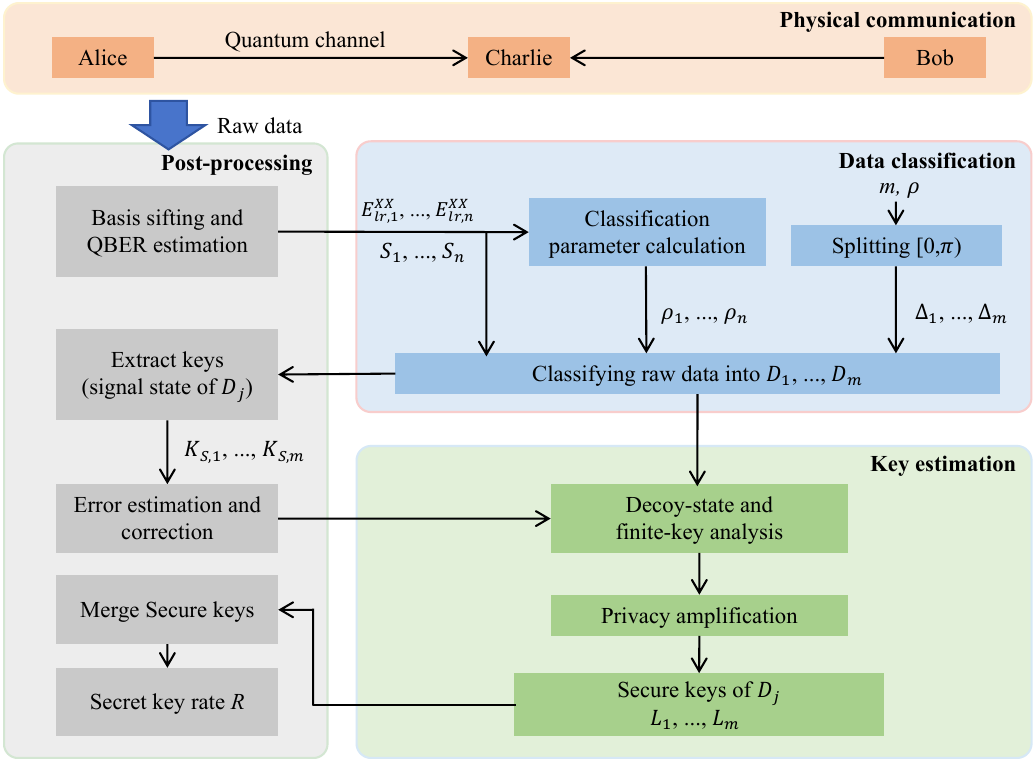}
		\caption{The schematic diagram of the proposed free-running RFI-MDI-QKD scheme. The scheme consists of four main stages: physical communication, data classification, key estimation, and post-processing. $S_i$ denotes the $i$-th subset obtained by dividing the total dataset according to the preset time interval $t$ and performing basis sifting, where $i=1,2,...,n$. $E_{lr,i}^{X X}$ represents the QBER corresponding to the intensity pair $lr$ selected by Alice and Bob in the $\mathit{XX}$ basis of subsets $S_{i}$. $\rho_{i}$ represents the classification parameter associated with the QBER of subset $S_i$. $m$ denotes the optimal number of partition intervals, and $\rho$ is the optimal initial point of the interval. $\Delta_j$ denotes the $j$-th interval obtained by partitioning $\left [ 0,\pi  \right )$ into $m$ intervals, where $j=1,2,...,m$. $D_j$ denotes the dataset composed of all subsets $S_i$ contained within the sub-interval $\Delta_j$. $K_{S,j}$ represents the raw key extracted from dataset $D_j$, and $L_{j}$ denotes the secure keys extracted from dataset $D_j$.}
		\label{Flowchart}
	\end{figure}
	
	In this study, we propose a free-running RFI-MDI-QKD scheme to counteract rapid reference-frame drift. By developing a classification distillation method that reclassifies total detection events based on varying quantum bit error rates (QBERs), our scheme can extract secure keys even under rapid reference-frame variations, without any changes to the experimental setup. We present a proof-of-principle experimental demonstration of the proposed scheme using a four-intensity joint-study RFI-MDI-QKD protocol operating at a 100 MHz repetition rate. Experimental results show that our system achieves a secret key rate (SKR) of 67.12 bit/s over a channel with 20 dB attenuation (including 5 km of optical fiber), nine times higher than that of the original scheme. When the reference-frame drift exceeds the threshold, our system generates an SKR of 16.43 bit/s over a channel with 24 dB attenuation, whereas the original scheme fails to generate any SKR. 	This work significantly enhances robustness against reference-frame drift, providing a promising solution for deploying quantum communication networks on mobile platforms.

	\begin{figure*}[t]
		\centering
		\includegraphics[width=0.9\linewidth]{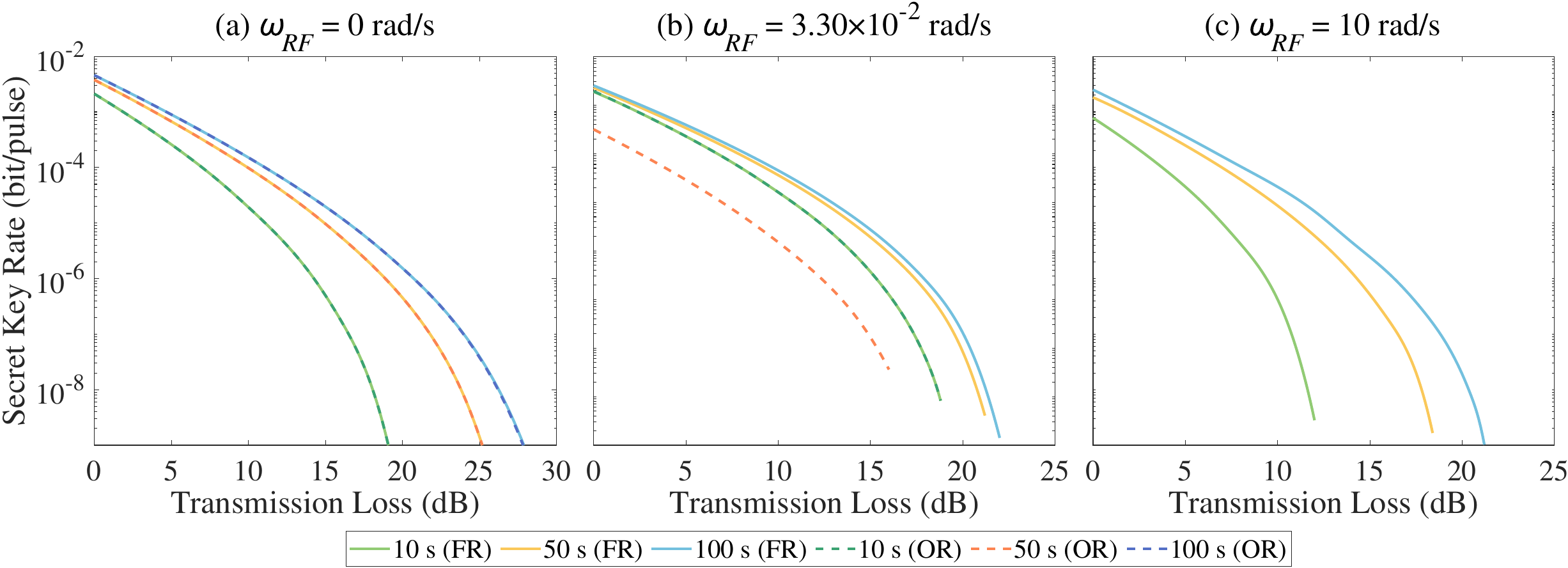}
				\caption{Performance comparison between the proposed free-running RFI-MDI-QKD (FR) scheme and the original RFI-MDI-QKD (OR) scheme under data-acquisition time ($T$) and reference-frame drift rates $(\omega_{RF})$. The solid and dashed lines represent the secret key rates of RF and OR schemes, respectively. In this simulation, both the dataset size $m $ and $\rho_{0}$ are optimized. (a) When $\omega _ {RF}=0 $ rad/s corresponds to the fixed reference frame, the secret key rate achieved by the proposed scheme is consistent with that of the original scheme. (b) When $\omega_{\mathrm{RF}} = 3.3\times 10^{-2}$ rad/s corresponds to laboratory reference-frame drift, the accumulated drift of the reference frame increasingly degrades the SKR of the original scheme as the data-acquisition time grows. In particular, once the acquisition time exceeds 50 s, the proposed scheme begins to exhibit significant advantages. (c) When $\omega_{\mathrm{RF}} = 10$ rad/s corresponds to actual buried optical fiber drift, the accumulated drift angle for all considered acquisition times surpasses the critical threshold, causing the original scheme to fail to generate any SKRs, whereas the proposed scheme still generates substantial SKRs. }
		\label{Simulation_drift}
	\end{figure*}
	
	\begin{figure*}[t]
		\centering
		\includegraphics[width=0.8\linewidth]{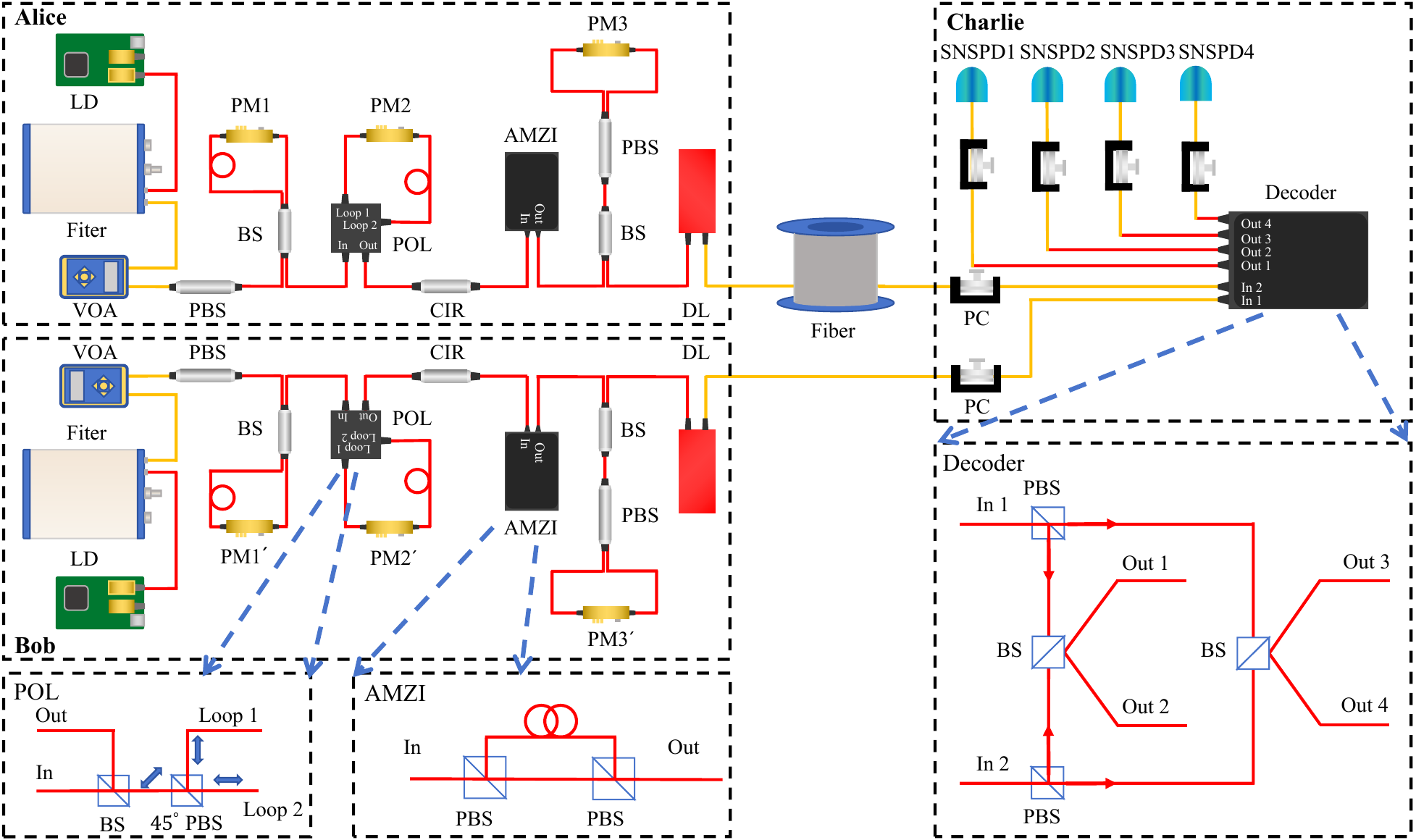}
		\caption{Experimental setup of RFI-MDI-QKD. LD, laser diode; Filter, bandwidth-variable tunable filter; VOA, variable optical attenuator; PBS, polarization beam splitter; BS, beam splitter; PM, phase modulator; POL, polarization modulator; $45^{\circ }$ PBS, polarizing beam splitter with $45^{\circ }$ rotation from the optical axis; CIR, circulator; AMZI, asymmetric Mach-Zehnder interferometer; DL, delay line; PC, polarization controller; SNSPD, superconducting nanowire single-photon detector.}
		\label{RFI-MDI-QKD}
	\end{figure*}
	
	To illustrate the proposed scheme, we adopt the well-known four-intensity RFI-MDI-QKD protocol as a reference (see Supplemental Material for details). As shown in Fig.~\ref{Flowchart}, state preparation and measurement follow the original RFI-MDI-QKD protocol, except that data classification and key estimation are integrated into a unified raw-data post-processing procedure.
	
	Specifically, during post-processing, all raw measurement data undergo basis sifting and are then divided into $n$ subsets according to a preset time interval $t$. This procedure yields $n$ subsets $S_i$, for which the corresponding quantum bit error rate (QBER) $E_{lr,i}^{XX}$ is calculated, with $i=1,2,\ldots,n$. The data then enter the classification stage, where a classification parameter $\rho_i$ is estimated for each subset using
	\begin{equation}
		\rho_{i}= \left\{\begin{array}{ll}
			\frac{1}{2}\arccos \!\left[\frac{\left(E_{lr,i}^{XX}-0.5\right)(y-1)(2y-x^{2}-2)}{(e_{d}^{XY}-0.5)x^{2}}\right], & E_{lr,i}^{XY} \ge 0.5, \\
			\pi-\frac{1}{2}\arccos \!\left[\frac{\left(E_{lr,i}^{XX}-0.5\right)(y-1)(2y-x^{2}-2)}{(e_{d}^{XY}-0.5)x^{2}}\right], & E_{lr,i}^{XY}<0.5 .
		\end{array}\right.
	\end{equation}
	Here, $x$ and $y$ are intermediate variables whose explicit expressions are provided in the Supplemental Material. The symbols $l$ and $r$ denote the intensities chosen by Alice and Bob, respectively, and $E_{lr,i}^{XY}$ represents the QBER for intensity pair $lr$ in the $\mathit{XY}$ basis of subset $S_i$. The parameter $e_{d}^{XY}$ denotes the intrinsic optical error rate in the $\mathit{X}$ or $\mathit{Y}$ basis.
	
	Next, an initial reference point $\rho$ is selected, and the phase interval $[0,\pi)$ is divided into $m$ slices $\Delta_j$, defined as
	\begin{equation}
		\Delta_j=\left\{\begin{array}{ll}
			\left\{\rho_i\,\big|\,\frac{m-1}{m}\pi+\rho\le\rho_i<\pi \;\text{or}\; 0\le\rho_i<\rho\right\}, & j=0,\\
			\left\{\rho_i\,\big|\,\frac{j-1}{m}\pi+\rho\le\rho_i<\frac{j}{m}\pi+\rho\right\}, & j>0,
		\end{array}\right.
	\end{equation}
	where $j=1,2,\ldots,m$, and $\rho$ is a constant that can be optimized to maximize the secret key rate. A subset $S_i$ is assigned to dataset $D_j$ if $\rho_i\in\Delta_j$.
	
	Each dataset $D_j$ is then independently processed in the post-processing and key-estimation stages. During post-processing, raw keys $\{K_{S,1},K_{S,2},\ldots,K_{S,m}\}$ are extracted from each dataset $D_j$, followed by error estimation and correction. In the key-estimation stage, decoy-state analysis and finite-key analysis are performed for each $D_j$, and the secure keys $L_j$ are obtained via privacy amplification. Finally, all secure keys are merged to compute the overall secret key rate,
	\begin{equation}
		R=\frac{L}{N_{\mathrm{tot}}}=\frac{\sum_{j=1}^{m}L_j}{N_{\mathrm{tot}}}
		=\frac{\sum_{j=1}^{m}N_jR_j}{N_{\mathrm{tot}}},
	\end{equation}
	where $N_{\mathrm{tot}}$ is the total number of pulses sent by Alice and Bob, and $N_j$ is the number of measurement events in dataset $D_j$. $R_j$ is  secret key rate of dataset $D_j$ and	detailed derivations of the secret key rate follow the standard four-intensity RFI-MDI-QKD protocol and are provided in the Supplemental Material.
	
	An intuitive explanation for the validity of the proposed scheme is given below. Under free-running reference frames, the original RFI theoretical framework predicts continuous accumulation of the misalignment angle $\beta$, leading to a monotonic increase in the QBERs of the $X$ and $Y$ bases (i.e., $E_{lr}^{XX}$, $E_{lr}^{XY}$, $E_{lr}^{YX}$, and $E_{lr}^{YY}$). Once the misalignment angle $\beta$ exceeds a critical threshold, all corresponding QBERs approach 50\%, rendering the protocol invalid and preventing secret key extraction~\cite{Zhang2014,Wang2016,Tang2022} (see Supplemental Material for a detailed analysis of the original protocol’s invalidation).
	
	In contrast, the proposed free-running RFI-MDI-QKD scheme classifies raw data acquired over extended time intervals during the data-classification stage. Consequently, by appropriately choosing $m$ and $\rho$, the QBERs $E_{lr}^{XX}$, $E_{lr}^{XY}$, $E_{lr}^{YX}$, and $E_{lr}^{YY}$ of the classified data can be prevented from simultaneously reaching 50\%.
	
	\itshape Simulation.\upshape—To demonstrate the applicability of the proposed scheme, we simulate the performance of the original scheme and the proposed scheme under the conditions of a fixed reference frame, laboratory reference-frame drift~\cite{Tang2022}, and actual buried optical fiber drift~\cite{Amies-King2023,Bersin2024,Maruyama2025}. As can been seen in Fig.~\ref{Simulation_drift}, the secret key rate (SKR) of the original RFI-MDI-QKD protocol is degrading  when  $\omega_{\mathrm{RF}} $ is rising from 0 rad/s to $3.3\times 10^{-2}$rad/s. Particularly, when  $\omega_{\mathrm{RF}}$ exceeding  10  rad/s, no key can be extract using the original scheme, whereas our proposed scheme still generates substantial SKRs.
	
	\itshape Setup.\upshape—The schematic of our experimental setup, shown in Fig.~\ref{RFI-MDI-QKD}, comprises three main parts: two transmitters (Alice and Bob), channel, and receiver (Charlie). Each transmitter has a gain-switched laser diode (LD) that generates phase-randomized laser pulses with a repetition rate of 100 MHz and a full width half maximum (FWHM) of about 140 ps. The generated light pulses are filtered through an ultra-narrow bandwidth-variable tunable filter to achieve a 3 dB bandwidth of approximately 31.8 pm, and are then attenuated to a single-photon level using a variable optical attenuator (VOA). Since the filter and VOA are single-mode devices, a polarizing beam splitter (PBS) is used to purify the polarization of the light pulses. Next, the light pulses are coupled into a Sagnac-based intensity modulator, which comprises a polarization-maintaining beam splitter (BS) and a phase modulator (PM). By adjusting the voltages applied to the PM1, the pulse intensities can be modulated to generate signal, decoy, and vacuum states for the decoy-state protocol.
	
	\begin{figure}[]
		\centering
		\includegraphics[width=1\linewidth]{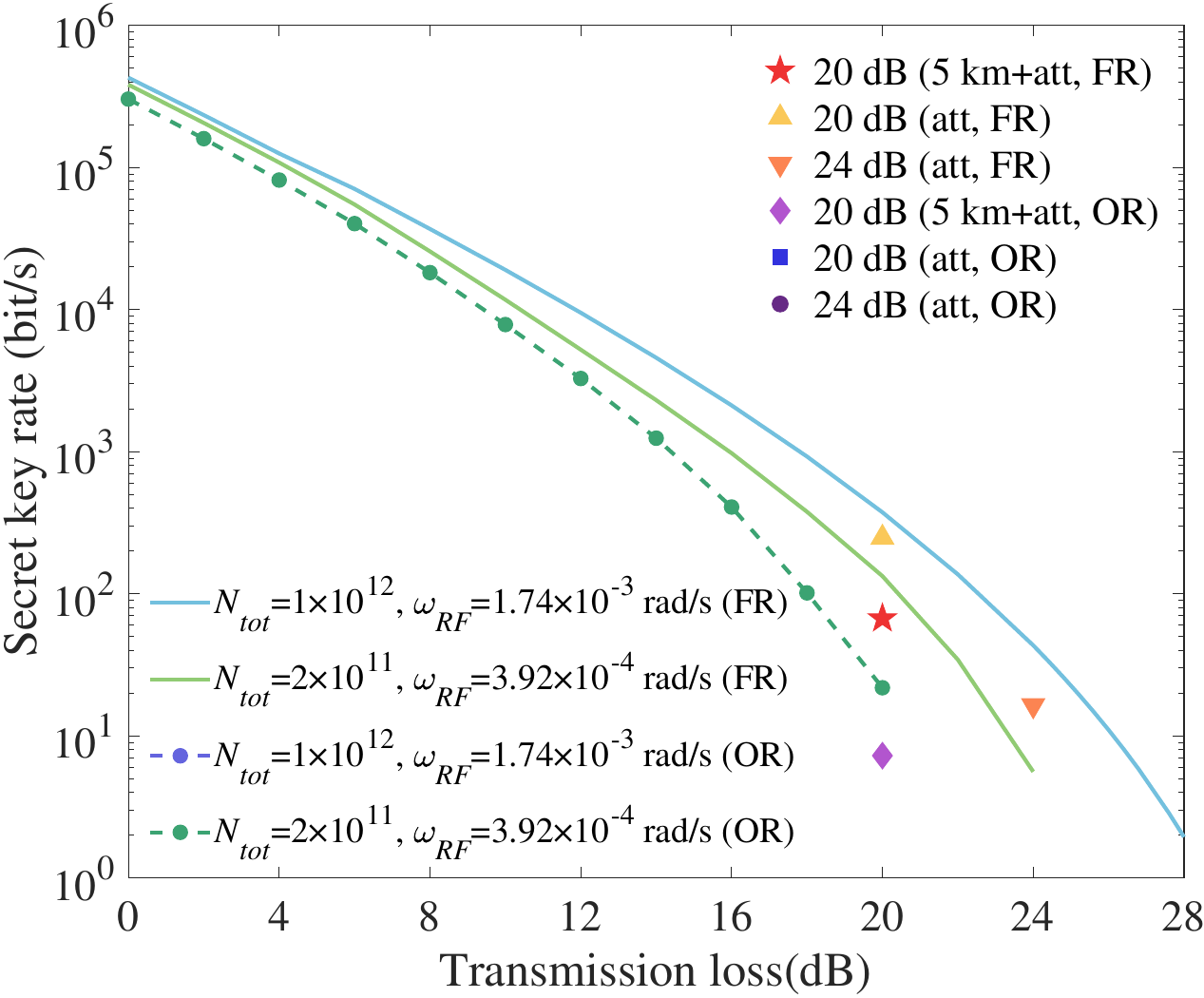}
		\caption{SKRs with different transmission loss. The solid and dashed lines represent the secret key rates of free-running (FR) and original (OR) schemes, respectively. The asterisks and diamonds denote experimental results of the FR and OR schemes with transmission loss of 20 dB (including 5 km fiber) at $N_{tot} = 2\times 10^{11}$ and $\omega_{RF} = 3.93\times 10^{-4}$ rad/s. The upright triangles and squares denote experimental results of the FR and OR schemes with transmission loss of 20 dB at $N_{tot} = 1\times 10^{12}$ and $\omega_{RF} = 1.60\times 10^{-3}$ rad/s. The inverted triangles and circles denote experimental results of the FR and OR schemes with transmission loss of 24 dB at $N_{tot} = 1\times 10^{12}$ and $\omega_{RF} = 3.14\times 10^{-4}$ rad/s.}
		\label{Key_Rate}
	\end{figure}
	
	The light pulses subsequently enter a Sagnac-based polarization modulator (POL), which comprises a BS, a $45^{\circ }$ PBS and a PM. It is used to modulate the four polarization states of $\left | H  \right \rangle$, $\left | +  \right \rangle$, $\left | V  \right \rangle$, and $\left | -  \right \rangle$ by applying phases $0$, $\frac{\pi }{2}$, $\pi$, and $\frac{3\pi }{2}$ to PM2, respectively. A circulator is used to prevent the light reflected back from the subsequent ring structure. The light pulses are then transmitted to an asymmetric Mach-Zehnder interferometer (AMZI) with a 2.5 ns delay, which splits the input $\left | H  \right \rangle$ polarization to the short ($S$) arm and the input $\left | V  \right \rangle$ polarization to the long ($L$) arm. When the input light pulses are $\left | +  \right \rangle$ or $\left | -  \right \rangle$ polarization state, the AMZI generates light pulses in both the $S$ and $L$ arms.
	
	Then, the light pulses are coupled into a phase modulation module, consisting of a BS, a PBS, and a PM, to generate states in $\mathit{X}$ or $\mathit{Y}$ basis. When the AMZI outputs light pulses in both the $L$ and $S$ arms, PM3 is utilized to apply an additional phase $\varphi _{\mathit{X}\left ( \mathit{Y} \right )} =0$ or $\pi$ ($\frac{\pi }{2}$ or $\frac{3\pi }{2}$) in the $L$ pulses to achieve $\mathit{X}$ ($\mathit{Y}$) basis modulation. Subsequently, the light pulses from Alice and Bob are transmitted to Charlie through a fiber channel, respectively. Delay lines (DLs) are used to precisely control the arrival timing of the light pulses from Alice and Bob at the receiver's BS.
	
	After the light pulses reach the receiver, two polarization controllers (PCs) actively compensate for polarization drift caused by the fiber channel of Alice and Bob, the compensation method is described in Supplementary Materials. Subsequently, two Bell states ($\left | \psi ^{+}  \right \rangle$ and $\left | \psi ^{-}  \right \rangle$) are measured in the decoder module. The light pulses in the two time-bins of polarization multiplexing are demultiplexed using PBSs and directed to BSs for interference. The total loss of receiver is approximately 1.40 dB.
	
	The interference results are detected by a superconducting nanowire single-photon detector (SNSPD; P-SPD8S, Photoec, China) with a detection efficiency of approximately 60\%, a dark count rate of about 20 Hz. The detection events are recorded using a time-to-digital converter (TDC) with a gate width of 800 ps. The HOM visibility of 44\% causes a 3\% optical intrinsic error in $\mathit{X}$ and $\mathit{Y}$ bases.

    \itshape Experimental results.\upshape—Figure~\ref{Key_Rate} presents the experimental results together with corresponding theoretical simulations. Channel attenuation is emulated using an optical attenuator, and the proposed scheme is experimentally verified at attenuation levels of 20 and 24~dB. For each attenuation setting, a total of $N_{\mathrm{tot}}=1\times10^{12}$ pulses are transmitted, after which the parameters $m$ and $\rho$ are optimized to maximize the secret key rate.

    Taking the 20~dB attenuation case as an example, the reference frame drifts by 16~rad over the entire 10\,000~s sampling duration, corresponding to an average drift rate of approximately $1.60\times10^{-3}$~rad/s. before data classification, The corresponding QBERs $E_{\omega\omega}^{XX}$, $E_{\omega\omega}^{XY}$, $E_{\omega\omega}^{YX}$, and $E_{\omega\omega}^{YY}$ are approximately 44.79\%, 46.81\%, 45.49\%, and 44.18\%, respectively, all close to 50\%, thereby preventing secret key generation.

    After data classification, the optimal initial point is $\rho=0.393$~rad, and the raw data are divided into four datasets ($m=4$). The resulting QBERs of the four datasets are (30.59\%, 47.89\%, 46.96\%, 30.24\%), (46.31\%, 33.31\%, 32.42\%, 44.43\%), (30.30\%, 47.29\%, 46.60\%, 30.35\%), and (48.20\%, 33.01\%, 33.66\%, 49.40\%), yielding extractable secret keys of $1.13\times10^{6}$, $1.16\times10^{6}$, $3.64\times10^{4}$, and $1.55\times10^{5}$ bits, respectively. The resulting total secret key rate is 248.14~bit/s. Similarly, at 24~dB attenuation, the achieved secret key rate is 16.43~bit/s, whereas the original scheme still fails to generate any secret key.

    \blk
	
	To further verify the practical applicability of our scheme, we introduce a 5 km fiber optic spool for experimental demonstration. Under 20 dB attenuation (including 5 km fiber), the total number of pulses sent by Alice and Bob is $N_{tot} = 2\times 10^{11}$, corresponding to reference frame drift rate of $\omega_{RF} = 3.93\times 10^{-4}$ rad/s. The achieved SKR is achieve an SKR of 67.12 bit/s, which is 9 times higher than that of the original scheme. Further detailed experimental results are summarized in Supplementary Materials. 
	
	\itshape Conclusions.\upshape—We propose a free-running RFI-MDI-QKD scheme to counteract rapid reference-frame drift, and validate its performance on an RFI-MDI-QKD system operating at a 100 MHz repetition rate. Simulation and experimental results demonstrate that the proposed scheme enables effective operation of the RFI-MDI-QKD system under rapidly varying reference frames. In particular, the proposed scheme can still generate substantial secure keys in scenarios where the original scheme fails to generate secure keys. Our scheme significantly enhances the robustness of the RFI-MDI-QKD protocol against reference-frame drift, facilitating the on-site deployment of quantum communication networks in complex environments. 

	\itshape Acknoledgement.\upshape—	The authors acknowledge B. Liu and S. Sun for the insightful discussions and B. Tang for providing theoretical assistance. 
	
	This study was supported by the National Natural Science Foundation of China (Nos. 62171144, 62031024, and  11865004), Guangxi Science Foundation (Nos.2021GXNSFAA220011, 2021AC19384 and 2025GXNSFAA069137), Open Fund of IPOC (BUPT) (No. IPOC2021A02), Guangdong Basic and Applied Basic Research Foundation (2024B1515120030), and Innovation Project of Guangxi Graduate Education (YCBZ2025064).

	\clearpage	
	
\end{document}